\documentstyle[aps,twocolumn,epsfig]{revtex}
\begin{document}
\draft
%%%%%%%%%%%%%%%%%%%%%%%%%%%%%%%%%%%%%%%%%%%%%%%%%%%
%%%%%%%%%%%% Begin Cover Page %%%%%%%%%%%%%%%%%%%%%%%%%%%%%%%%%%%
\preprint{\begin{minipage}[b]{1.5in}
          UK/TP 99-18\\
          hep-ph/9911476\\
          \end{minipage}}
\vspace{0.2in}

\title{Universality in nuclear dependence coefficient 
$\alpha(q_T)$}
\author{Xiaofeng Guo$^1$, Jianwei Qiu$^2$, and Xiaofei Zhang$^2$}
\address{$^1$Department of Physics and Astronomy,
             University of Kentucky \\
             Lexington, Kentucky 40506, USA\\
         $^2$Department of Physics and Astronomy,
             Iowa State University \\
             Ames, Iowa 50011, USA }
\date{February 16, 2000}
\maketitle

\begin{abstract}
We derive the nuclear dependence coefficient $\alpha(q_T)$
for Drell-Yan and J/$\psi$ production.  We show that at small $q_T$, 
the $\alpha(q_T)$ is given by an universal functional form:\
$\alpha(q_T)=a+b\, q_T^2$, and the parameters $a$ and $b$ are
completely determined by either perturbatively calculable or
independently measurable quantities. This universal functional form
$\alpha(q_T)$ is insensitive to the $A$, and 
is consistent with existing data. 
\end{abstract}
\vspace{0.2in}

Anomalous nuclear dependence in transverse momentum spectrum of  
single hadron production in high energy hadron-nucleus collisions,
known as Cronin effect~\cite{Cronin}, was first observed nearly 25
years ago. In recent years, new data on nuclear dependence for 
the Drell-Yan \cite{NA10,E772,Eloss} and J/$\psi$
production~\cite{E866,Jpsi-Et,NA3} 
have renewed our interests in the observed novel effect
\cite{alpha-thy}.  However, due to technical difficulties in
handling multi-scale calculations,
QCD predictions for such  nuclear dependence only exist
for large transverse momentum ($q_T$) region 
\cite{LQS1,Guo1,BMuller},
while almost all existing data are in small $q_T$ region 
\cite{NA10,E772,Eloss,E866,Jpsi-Et,NA3}.  In this letter, 
we derive the nuclear dependence coefficient $\alpha(q_T)$ for
Drell-Yan and J/$\psi$ production in small $q_T$ region.

For the Drell-Yan production, the $\alpha(q_T)$ is defined as  
\begin{equation}
\frac{d\sigma^{hA}}{dQ^2 dq_T^2} =
A^{\alpha(q_T)} \times \frac{d\sigma^{hN}}{dQ^2 dq_T^2} \ , 
\label{alpha}   
\end{equation}  
where $Q$ is the Drell-Yan pair's total invariant mass,
$A$ is the atomic weight of the target, 
and $d\sigma^{hA}/dQ^2 dq_T^2$ and 
$d\sigma^{hN}/dQ^2 dq_T^2$ are the transverse momentum spectrum
in hadron-nucleus and hadron-nucleon collisions, respectively. 
We argue below that in small $q_T$ region ($q_T<2.5$~GeV), the
Drell-Yan $q_T$-spectrum at fixed target energies 
in {\it both} hadron-hadron and hadron-nucleus collisions 
can be well represented by a
Gaussian-like distribution, and corresponding parameters can be 
completely determined by the quantities that are either calculable in 
QCD perturbation theory or independently measurable in experiments.   
From such a Gaussian-like spectrum,
we derive our universal functional form for the $\alpha(q_T)$.

Depending on physics origins, the Drell-Yan $q_T$-spectrum 
can be divided into three regions: (I) small
$q_T$ region where  $q_T< q_T^S \sim 1$~GeV; (II)  
intermediate $q_T$ region where $q_T^S \leq q_T \leq
q_T^L$ with $q_T^L \equiv \kappa Q$ and $\kappa\sim 1/3-1/2$; and 
(III) large $q_T$ region where $q_T>q_T^L$.  In region (I), the
Drell-Yan $q_T$-spectrum 
in hadron-nucleon collisions is dominated by the
intrinsic transverse momenta of colliding partons.  In perturbative
calculations, the spectrum at the leading order is proportional to
$\delta^2(\vec{q}_T)$.  If the partons' intrinsic transverse momentum
has a Gaussian-like distribution, its effects 
can be included by replacing the $\delta$-function with 
$\delta(q_x)={\rm lim}_{\tau\rightarrow 0}\,
\frac{1}{\sqrt{2\pi}\,\tau}\, {\rm exp}[-q_x^2/2\tau^2]$, and we
obtain the Drell-Yan $q_T$-spectrum 
in region (I):  
\begin{equation}
\frac{d\sigma^{(I)}}{dQ^2 dq_T^2} = N_{DY}\, \frac{1}{2\tau^2}\,
{\rm e}^{-q_T^2/2\tau^2}\, , 
\label{DYGau}
\end{equation}
with $N_{DY}$ a dimensional normalization and $\tau$ the 
width.  

In region (II),  the spectrum can be calculated in QCD
perturbation theory with resummation of large logarithms, such as
$(\alpha_s\ln^2 (Q^2/q_T^2)) ^n$, which are due to the gluon radiations 
from incoming partons \cite{Resum}.  The resummation is 
extremely important for
$W^{\pm}$ and $Z^0$ production at collider energies because the 
$Q^2/q_T^2$ can be as large as $8\times 10^{3}$ for $q_T\sim
1$~GeV.  However, for the Drell-Yan production at fixed target energies,
the resummation is much less important because of much smaller value
of $Q^2/q_T^2$.  In fact, as shown in Ref.~\cite{GQZ}, all existing
data for $q_T$ as large as $2.5$~GeV can be well represented by
extending the Gaussian-like distribution in Eq.~(\ref{DYGau}) from 
region (I) to region (II). 

In region (III),  the transverse momentum spectrum can be 
calculated in perturbative QCD \cite{Berger}.  
Therefore, the Drell-Yan $q_T$-spectrum at
fixed target energies can be represented by 
\begin{equation}
\frac{d\sigma}{dQ^2dq_T^2} = 
  \frac{d\sigma^{(I)}}{dQ^2dq_T^2}
+ \left[\frac{d\sigma^{(III)}}{dQ^2dq_T^2}
       -\frac{d\sigma^{(I)}}{dQ^2dq_T^2} \right] \theta(q_T-q_T^L),
\label{DYqt}
\end{equation}
where $d\sigma^{(III)}/dQ^2dq_T^2$ is the perturbatively
calculated $q_T$-spectrum for $q_T>q_T^L$, and
$d\sigma^{(I)}/dQ^2dq_T^2$, defined in Eq.~(\ref{DYGau}), fits data in
regions (I)+(II).  

Using moments of the Drell-Yan $q_T$-spectrum in Eq.~(\ref{DYqt}), we 
can relate  $N_{DY}$ and $\tau$ in Eq.~(\ref{DYGau}) to 
physical quantities.  At fixed target energies,
the contribution from the second term in Eq.~(\ref{DYqt}) 
to $d\sigma/dQ^2$ is much less than {\it one} percent.  
Therefore, up to  less than one
percent uncertainty, $N_{DY}\approx d\sigma/dQ^2$.

Define the averaged transverse momentum square as 
$\langle q_T^2 \rangle \equiv \int dq_T^2\, q_T^2\, 
(d\sigma/dQ^2dq_T^2) / (d\sigma/dQ^2)$.  We find \cite{GQZ} that 
the contribution of the second term in Eq.~(\ref{DYqt}) to 
$\langle q_T^2 \rangle$ is much less than {\it ten} percent of 
the first term.  Therefore, by iteration, we obtain 
$2\tau^2 \approx \langle q_T^2 \rangle - \Gamma(q_T^L)$ with 
\begin{equation}
\Gamma(q_T^L)\equiv 
\frac{1}{d\sigma/dQ^2} \int_{q_T^L}\, q_T^2 \left[
    \frac{d\sigma^{(III)}}{dQ^2dq_T^2} 
  - \frac{d\sigma^{(I)}}{dQ^2dq_T^2} \right] dq_T^2,
\label{Gamma-def} 
\end{equation}
where $2\tau^2$ in $d\sigma^{(I)}/dQ^2dq_T^2$ in Eq.~(\ref{Gamma-def})
is approximately given by $\langle q_T^2 \rangle$.  
Substituting the $N_{DY}$ and $\tau$ into
Eq.~(\ref{DYGau}), we obtain the Drell-Yan spectrum for $q_T<q_T^L$ 
as 
\begin{equation}
\frac{d\sigma^{hN}}{dQ^2 dq_T^2} =
\frac{d\sigma^{hN}/dQ^2}
     {\langle q_T^2\rangle^{hN}-\Gamma(q_T^L)^{hN}} \,
{\rm e}^{-q_T^2/ (\langle q_T^2 \rangle^{hN}-\Gamma(q_T^L)^{hN})}.
\label{qtexpN}
\end{equation}   
Most importantly, the $\Gamma(q_T^L)$ in Eq.~(\ref{qtexpN})
is small and perturbatively calculable.

For the Drell-Yan $q_T$-spectrum  in  hadron-nucleus collisions, 
 we also need to consider multiple scattering.
Similar to the single-scattering case, at
the leading order in perturbation theory, the double-scattering
contribution is also proportional to a
$\delta$-function \cite{Guo2},    
\begin{equation}
\frac{d\sigma^{hA}_D}{dQ^2 d^2q_T}
\propto T_{qg}(x,x_1,x_2,k_T) \, \delta^2(\vec{q}_T-\vec{k}_T)\ ,
\label{DO-spectrum}
\end{equation}
where the subscript $D$ indicates the double scattering,
$T_{qg}(x, x_1,x_2,k_T)$ is the quark-gluon correlation function
\cite{Guo2,LQS2}, where $x,x_1$, and $x_2$ are the momentum fractions
carried by the quark and gluon fields.  The $k_T$ in
Eq.~(\ref{DO-spectrum}) represents the intrinsic momentum of the gluon
which gives additional scattering.  Following the same arguments
leading to Eq.~(\ref{qtexpN}), we can show \cite{GQZ} that if the
partons' intrinsic $k_T$-dependence has a Gaussian-like distribution, 
the double scattering contributions to the Drell-Yan $q_T$-spectrum in
small $q_T$ region can also be represented by a Gaussian form.

In high $q_T$ region, the Drell-Yan $q_T$-spectrum in
hadron-nucleus collision also has a perturbative tail. The  
Drell-Yan $q_T$-spectrum at large $q_T$ in
hadron-nucleus collisions was calculated in Ref.~\cite{Guo1}.
The nuclear dependence of Drell-Yan $q_T$-spectrum depends on two 
types of multiparton correlation functions inside the nucleus: 
$T^{DH}$ and $T^{SH}$, which correspond to the double-hard and 
soft-hard double scattering subprocesses respectively \cite{Guo1}.  
These correlation functions are as fundamental as the
well-known parton distributions, and can be extracted from other
physical observables.  

Similar to  deriving  Eq.~(\ref{qtexpN}), we can derive
\begin{equation}
\frac{d\sigma^{hA}}{dQ^2 dq_T^2} = 
\frac{d\sigma^{hA}/dQ^2}
     {\langle q_T^2\rangle^{hA}-\Gamma(q_T^L)^{hA}} \,
{\rm e}^{-q_T^2/ (\langle q_T^2 \rangle^{hA}-\Gamma(q_T^L)^{hA})},
\label{qtexpA}
\end{equation}   
where $\langle q_T^2 \rangle^{hA} = 
\langle q_T^2 \rangle^{hN} + \Delta \langle q_T^2\rangle^{hA}$. 
$\Delta\langle q_T^2 \rangle^{hA}$ is the
transverse momentum broadening and calculable in QCD perturbation
theory \cite{Guo2,LQS2}. In Eq.~(\ref{qtexpA}), $\Gamma(q_T^L)^{hA}$ 
is a small
contribution to $\langle q_T^2 \rangle^{hA}$, and it is calculable and
depends on the perturbative tail of the $q_T$-spectrum.

Substituting Eqs.~(\ref{qtexpN}) and (\ref{qtexpA}) into 
Eq.~(\ref{alpha}), we derive  $\alpha(q_T)$ for the Drell-Yan
production in small $q_T$ region:
\begin{eqnarray}
\alpha_{DY}(q_T) &=& 1+\, 
\frac{1}{\ln(A)} \left[ \ln\left(R^A_{DY}(Q^2)\right) 
+ \ln\left(\frac{1}{1+\chi_{DY}}\right)  \right. 
\nonumber \\ 
&\ & {\hskip 0.4in} \left. 
 + \frac{\chi_{DY}}{1+\chi_{DY}}\, 
\frac{q_T^2}{\langle q_T^2 \rangle^{hN}-\Gamma(q_T^L)^{hN}} 
\right]\, ,
\label{alpha-qt}
\end{eqnarray}
where $R^A_{DY}(Q^2)\equiv (1/A)(d\sigma^{hA}/dQ^2) / 
(d\sigma^{hN}/dQ^2)$.  The $\chi_{DY}$ in 
Eq.~(\ref{alpha-qt}) is defined by
\begin{equation}
\chi_{DY} \equiv 
\frac{\Delta \langle q_T^2 \rangle^{hA}-\Delta\Gamma(q_T^L)^{hA}} 
     {\langle q_T^2 \rangle^{hN}-\Gamma(q_T^L)^{hN}} 
\approx 
\frac{\Delta \langle q_T^2 \rangle^{hA}}{\langle q_T^2 \rangle^{hN}},
\label{chiDY}
\end{equation}
where $\Delta\Gamma(q_T^L)^{hA}\equiv \Gamma(q_T^L)^{hA} -
\Gamma(q_T^L)^{hN}$, and is  much smaller than
$\Delta\langle q_T^2 \rangle^{hA}$ \cite{GQZ}.
The $\alpha_{DY}(q_T)$ in Eq.~(\ref{alpha-qt}) has a
quadratic dependence on $q_T$. 

At the leading order in $\alpha_s$, 
$\Delta\langle q_T^2\rangle^{hA} \propto T^{SH}_{qg}$ \cite{Guo2}. 
For normal nuclear
targets, $T^{SH}_{qg}(x)=\lambda^2
A^{4/3} q(x)$ with $q(x)$ the normal quark distribution and the
parameter $\lambda^2$ proportional to the size of averaged color field
strength square inside a nuclear target \cite{LQS1}.  
Consequently, we have  $\Delta\langle q_T^2\rangle^{hA} = b_{DY}\,
A^{1/3}$  with $b_{DY} \propto \lambda^2$.
In this letter, we use a measured $b_{DY}\approx 0.022$~GeV$^2$
\cite{DY-b} to fix $\lambda^2$ and $T^{SH}$.
Taking the small $\chi_{DY}$ limit, and using the fact that 
$R^A_{DY}(Q^2) \approx 1$, we derive
\begin{eqnarray}
\alpha_{DY}(q_T) & \approx &
1+\, \frac{b_{DY}}{\langle q_T^2 \rangle^{hN}}\, 
\left[-1 + \frac{q_T^2}{\langle q_T^2 \rangle^{hN}} \right] \, .
\label{alpha-qt0} 
\end{eqnarray}
In deriving Eq.~(\ref{alpha-qt0}), we used $A^{1/3}\sim \ln(A)$, which
is a good approximation for most relevant targets.
Eq.~(\ref{alpha-qt0}) shows that the leading contribution 
to $\alpha_{DY}(q_T)$ does not depend on the $A$.
\begin{figure}
\epsfig{figure=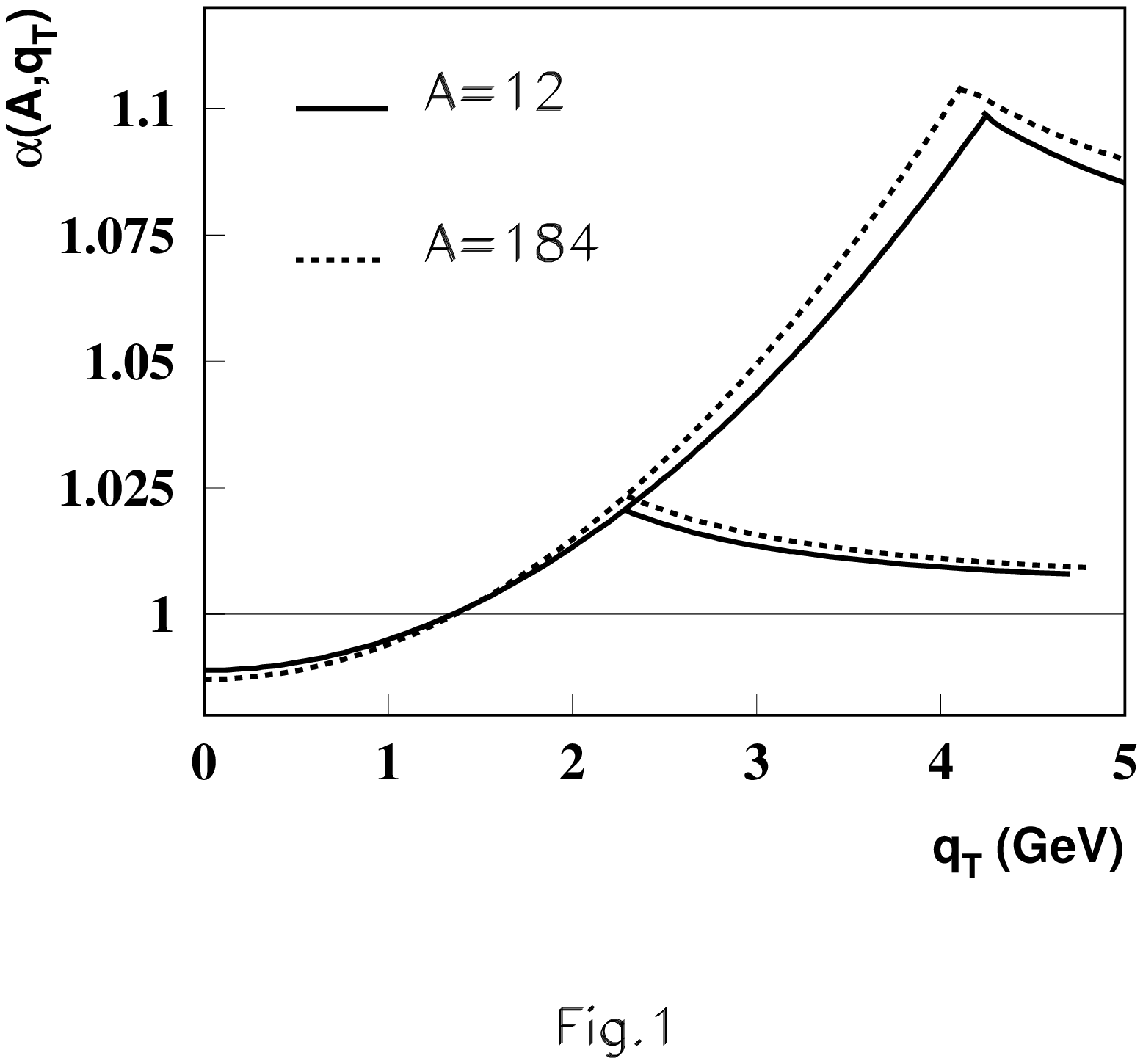,width=1.0in}
\caption{$\alpha_{DY}(q_T)$ for the Drell-Yan production as a function
of $q_T$.  At small $q_T$, the $\alpha_{DY}(q_T)$ in
Eq.~(\protect\ref{alpha-qt}) is used. At large $q_T$, QCD predictions 
from Ref.~\protect\cite{Guo1} are plotted with 
$r_0 = 1.1$~fm and $[xG(x)]_{x\approx 0}=3$ 
for the possible maximum and minimum values of the $C$.}  
\label{fig1}
\end{figure} 

\begin{figure}
\epsfig{figure=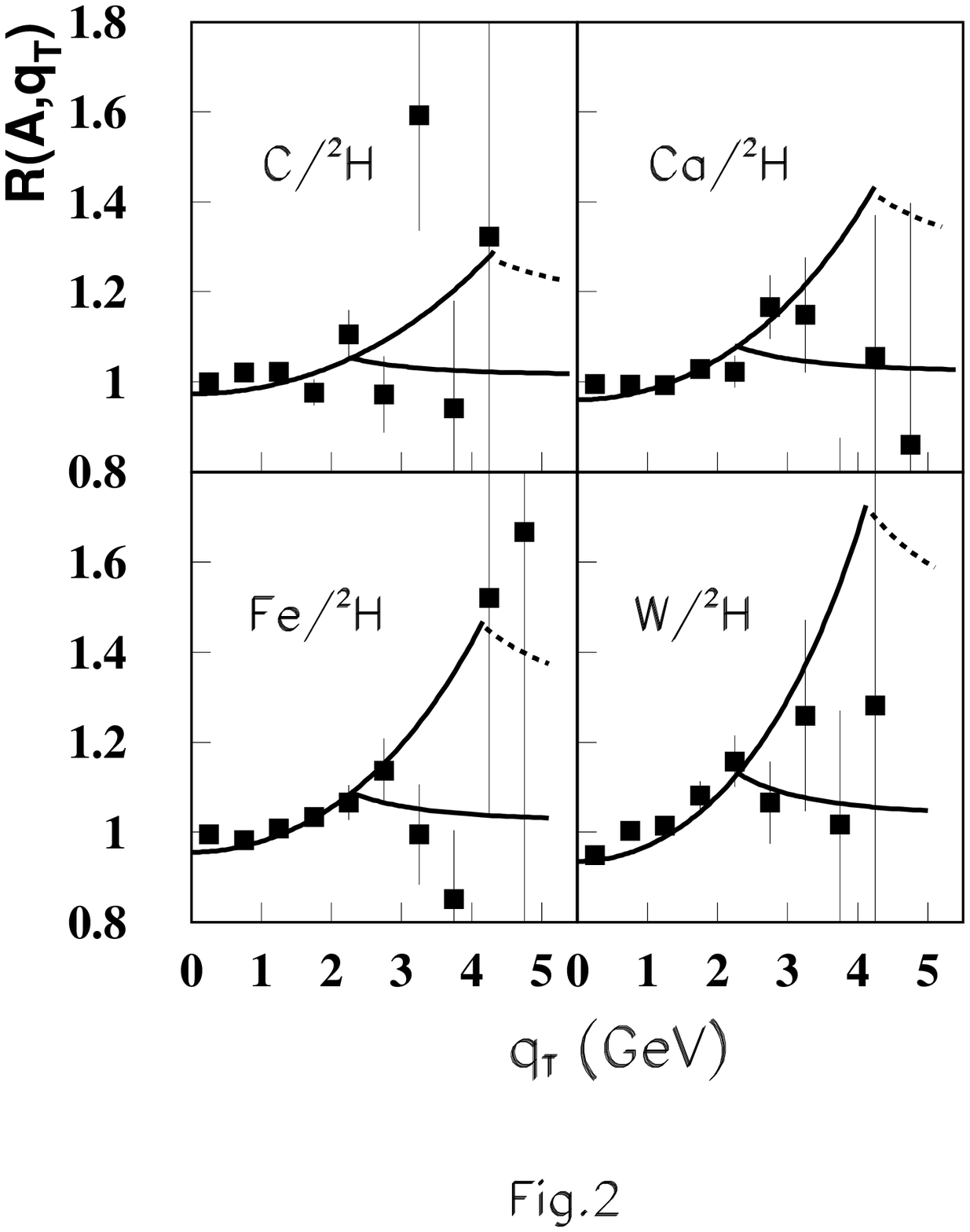,width=1.0in}
\caption{Comparison of our $\alpha_{DY}(q_T)$ in 
Eq.~(\protect\ref{alpha-qt}) with the Drell-Yan data from Fermilab
E772 collaboration \protect\cite{E772,E772-moss}.}
\label{fig2}
\end{figure}

In Fig.~\ref{fig1}, we plot $\alpha_{DY}(q_T)$ 
 for targets with $A=12$ and 
$A=184$.  At small $q_T$, we used 
Eq.~(\ref{alpha-qt}) and set the small $\Gamma(q_T^L)=0$. 
We also used the measured 
$\langle q_T^2\rangle^{hN} = 1.8$~GeV$^2$, and 
$\Delta\langle q_T^2\rangle^{hA} =0.022 A^{1/3}$~GeV$^2$ 
\cite{GQZ,DY-b}.  
At large $q_T$, we plot the QCD predictions from
Ref.~\cite{Guo1}, which depend on both $T^{SH}$ and $T^{DH}$.  
$T^{SH}$ is fixed by  $\Delta\langle q_T^2\rangle^{hA}$. 
However, there is no direct observable yet to extract
$T^{DH}$ \cite{BMuller}.  Because of 
the operator definition of  $T^{DH}(x_1,x_2)$, it was
 assumed that 
$T^{DH}_{f_1 f_2}(x_1,x_2)=(2\pi C) A^{4/3}\,f_1(x_1)\,f_2(x_2)$ with
$f_1$ and $f_2$ parton distributions of flavor $f_1$ and $f_2$
\cite{Guo1}.  
Assuming no  quantum interference between different nucleon states, 
one derives $C=0.35/(8\pi r_0^2)$~GeV$^2$ 
with $r_0 \approx 1.1 -1.25$~fm, which is just a geometric factor 
for finding two nucleons at 
the same impact parameter \cite{Guo1}.  On the other hand, when $x_1$
(or $x_2$) goes to zero, the corresponding parton fields reach the 
saturation region, and the $T^{DH}$ is reduced to $T^{SH}$. Therefore,
we have $C\approx \lambda^2/(2\pi [xG(x)]_{x\approx 0})$ \cite{GQZ}, 
where $[xG(x)]_{x\approx 0}$ is of the order of unity \cite{Mueller}.
Because of a combination of a small value of the measured $\lambda^2$
from Drell-Yan data \cite{DY-b} and a choice of 
$[xG(x)]_{x\approx 0} \approx 3$ \cite{GQZ}, 
these two approaches result into a factor of 20
difference in numerical value for the parameter $C$ \cite{GQZ}.  The  
value for the $C$ obtained in Ref.~\cite{Guo1} without any 
quantum interference should represent a possible maximum for $C$, 
while the value obtained in Ref.~\cite{GQZ} with full quantum 
interference (in saturation region) 
represents a possible minimum for the $C$.
In Fig.~\ref{fig1}, we plot the perturbatively calculated
$\alpha_{DY}(q_T)$ \cite{Guo1} at large $q_T$  
with the maximum and minimum values of the $C$
discussed above, and let the $\alpha(q_T)$ in small $q_T$ naturally
linked to that at large $q_T$.  
Fig.~\ref{fig1} also shows that $\alpha_{DY}(q_T)$   
is insensitive to the   
atomic number $A$. In Fig.~\ref{fig2},  we plot $R(A,q_T)\equiv
A^{\alpha_{DY}(q_T)}$ by using $\alpha_{DY}(q_T)$ in Fig.~\ref{fig1} 
and compare our predictions with 
data from E772 \cite{E772,E772-moss}. Without any extra free fitting
parameters, our predictions shown in Fig.~\ref{fig2} are consistent
with data at small $q_T$, and due to the large error in data
at high $q_T$, current Drell-Yan data  are
consistent with almost any value for the $C$ between 
the maximum and the
minimum discussed above.  In order to test the theory and pin
down the value of the $C$, we need either better data or different
observables.  Because of the strong dependence on the double-hard
subprocesses, the angular dependence of the Drell-Yan pair 
\cite{BMuller} could be an excellent signal for measuring the $C$.

Similarly, we can also obtain the 
$\alpha(q_T)$ for J/$\psi$ production. 
Kinematically, hadronic J/$\psi$ production is  like
Drell-Yan production with $Q\sim M_{{\rm J/}\psi}$, and  
its $q_T$-spectrum can also be characterized by 
three different regions as in the Drell-Yan case.
Because J/$\psi$ mass $M_{{\rm J/}\psi}$ is smaller than any typical
$Q$ measured for the Drell-Yan continuum, the logarithm $(\alpha_s
\ln^2(M_{{\rm J/}\psi}/q_T^2))^n$ for J/$\psi$ production is 
less important. Consequently, at fixed target energies, 
a Gaussian-like distribution can fit
J/$\psi$'s $q_T$-spectrum even better.  
Therefore, following the same arguments used above for the Drell-Yan
production, we derive $\alpha(q_T)$
for J/$\psi$ production in small $q_T$ region,
\begin{eqnarray}
\alpha_{{\rm J/}\psi}(q_T) &=& 1+\, 
\frac{1}{\ln(A)} \left[ \ln\left(R^A_{{\rm J/}\psi}\right) 
+ \ln\left(\frac{1}{1+\chi_{{\rm J/}\psi}}\right) \right. 
\nonumber \\ 
&\ & {\hskip 0.2in} \left. 
 + \frac{\chi_{{\rm J/}\psi}}{1+\chi_{{\rm J/}\psi}}\, 
\frac{q_T^2}{\langle q_T^2 \rangle^{hN}_{{\rm J/}\psi} 
               -\Gamma(q_T^L)^{hN}_{{\rm J/}\psi}} 
\right],
\label{a-JPsi}
\end{eqnarray}
where $R^A_{{\rm J/}\psi}\equiv (1/A)\sigma^{hA}_{{\rm J/}\psi}
/\sigma^{hN}_{{\rm J/}\psi}$, and 
$\chi_{{\rm J/}\psi}$ is defined by 
\begin{equation}
\chi_{{\rm J/}\psi} 
\equiv 
\frac{\Delta \langle q_T^2 \rangle^{hA}_{{\rm J/}\psi} 
     -\Delta\Gamma(q_T^L)^{hA}_{{\rm J/}\psi}} 
     {\langle q_T^2 \rangle^{hN}_{{\rm J/}\psi} 
     -\Gamma(q_T^L)^{hN}_{{\rm J/}\psi}} 
\approx 
\frac{\Delta \langle q_T^2 \rangle^{hA}_{{\rm J/}\psi}}
     {\langle q_T^2 \rangle^{hN}_{{\rm J/}\psi}}.
\label{chiJPsi}
\end{equation}
Similar to the Drell-Yan case, $\Gamma(q_T^L)^{hN}_{{\rm J/}\psi}$ and 
$\Delta\Gamma(q_T^L)^{hN}_{{\rm J/}\psi}$ are perturbatively 
calculable, and much smaller
than $\langle q_T^2 \rangle^{hN}_{{\rm J/}\psi}$ and 
$\Delta\langle q_T^2 \rangle^{hN}_{{\rm J/}\psi}$,
respectively \cite{GQZ}. 

One major difference between J/$\psi$ production and the 
Drell-Yan process is the nuclear dependence of 
$R^A_{{\rm J/}\psi}$.  Clear nuclear suppression for 
$\sigma^{hA}_{{\rm J/}\psi}$ has been observed \cite{JPsi-sup}.  
Since we are only interested in the general features of
$\alpha(q_T)$, we adopt the following 
simple parameterization
\begin{equation}
R^A_{{\rm J/}\psi} = \frac{1}{A}\, 
\frac{\sigma_{{\rm J/}\psi}^{hA}}{\sigma_{{\rm J/}\psi}^{hN}}
 = {\rm e}^{-\beta A^{1/3}},
\label{JPsiA}
\end{equation}
which fits all experimental data on J/$\psi$ suppression in
hadron-nucleus collisions \cite{BQV,Dima}.  The $A^{1/3}$ factor in
Eq.~(\ref{JPsiA}) represents an effective medium length, and the
$\beta$ is a constant \cite{GQZ}.

Similar to the Drell-Yan process,  
$\Delta\langle q_T^2\rangle^{hA}_{{\rm J/}\psi}$ 
is proportional to four-parton correlation functions \cite{GQZ}. 
Due to final-state interactions for J/$\psi$ production, 
$\Delta\langle q_T^2\rangle^{hA}_{{\rm J/}\psi}$
depend on both quark-gluon and gluon-gluon correlation functions.
We can define \cite{GQZ} that 
$\Delta\langle q_T^2\rangle^{hA}_{{\rm J/}\psi} 
= b_{{\rm J/}\psi}\, A^{1/3}$, 
where $b_{{\rm J/}\psi}$ can be calculated or extracted from 
data.  From Ref.~\cite{DY-b}, we obtain $b_{{\rm J/}\psi}\approx
0.06$~GeV$^2$.  With 
$\langle q_T^2 \rangle^{hN}_{{\rm J/}\psi}
\approx 1.68$~GeV$^2$ \cite{GQZ}, we can also take the
small $\chi_{{\rm J/}\psi}$ limit in Eq.~(\ref{a-JPsi}) and  obtain
\begin{equation}
\alpha_{{\rm J/}\psi}(q_T) \approx  
1 - \beta 
+ \frac{b_{{\rm J/}\psi}}
       {\langle q_T^2\rangle^{hN}_{{\rm J/}\psi}}
\left[ -1 + \frac{q_T^2}{\langle q_T^2 \rangle^{hN}_{{\rm J/}\psi}} 
\right]  ,
\label{a-JPsi0}
\end{equation}
where $A^{1/3}\sim \ln(A)$ was again used.
It is clear from Eq.~(\ref{a-JPsi0}) that in small $q_T$ region, 
$\alpha_{{\rm J/}\psi}(q_T)$ is also insensitive to the atomic number 
$A$ of targets.  
Furthermore, the nuclear suppression in $R^A_{{\rm J/}\psi}$ 
corresponds to a $q_T$-independent shift in the magnitude of 
$\alpha_{{\rm J/}\psi}(q_T)$.

\begin{figure}
\epsfig{figure=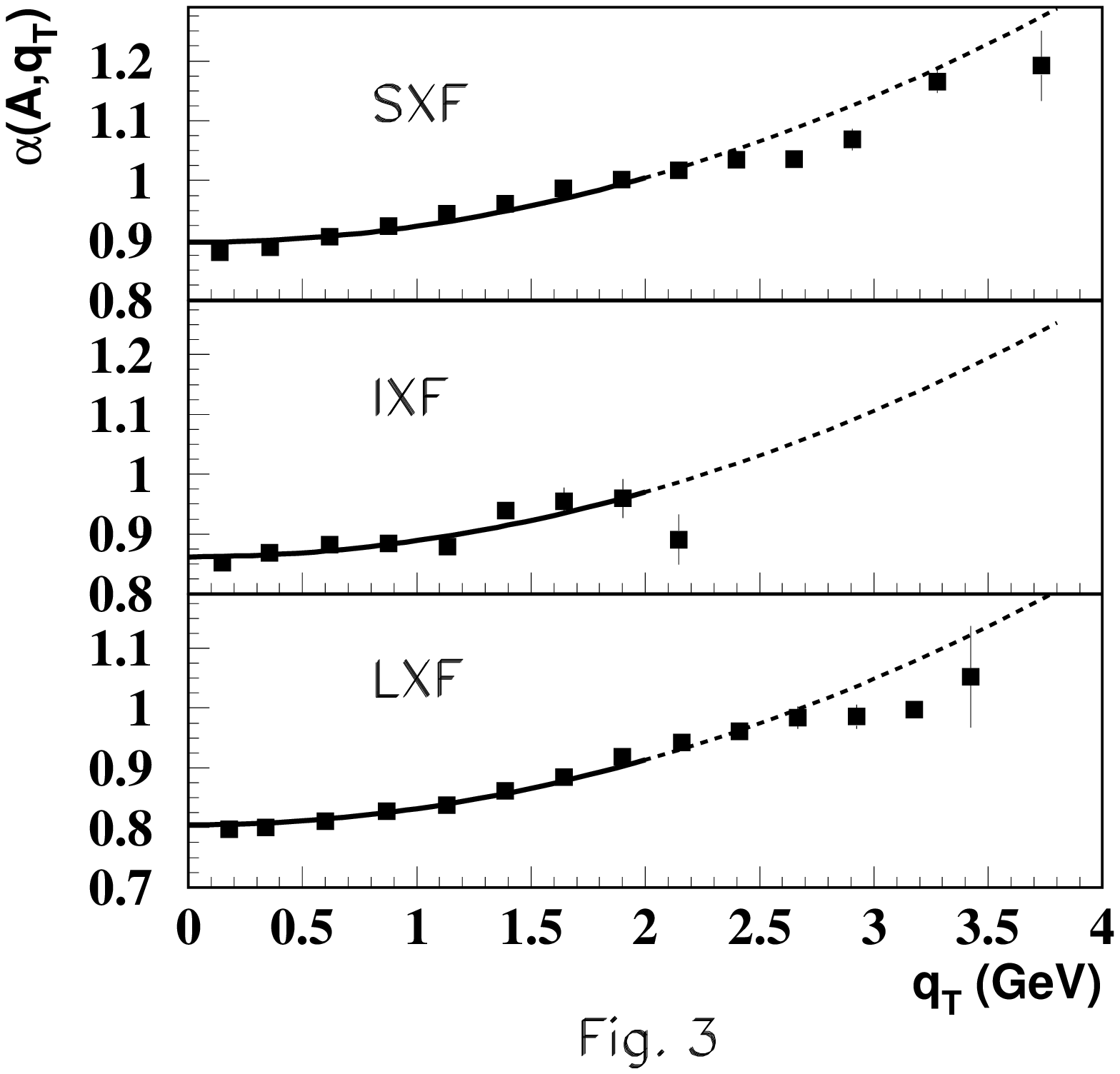,width=1.0in}
\caption{Comparison of our fits, using the $\alpha_{{\rm
J/}\psi}(q_T)$ in Eq.~(\protect\ref{a-JPsi}), with J/$\psi$ data from 
Fermilab E866 collaboration \protect\cite{E866} in three different 
regions of $x_F$.}
\label{fig3}
\end{figure}

Because the $R^A_{{\rm J/}\psi}$, $\chi_{{\rm J/}\psi}$, and other
physical quantities in Eq.~(\ref{a-JPsi}) can depend on $x_F$, 
$\alpha_{{\rm J/}\psi}(q_T)$ can also be a function of $x_F$.
Experiments show that the larger 
$x_F$, the more suppression for J/$\psi$ production (or smaller
$R^A_{{\rm J/}\psi}$) \cite{JPsi-xf-exp}. Consequently, from 
Eq.~(\ref{a-JPsi}), we will have smaller 
$\alpha_{{\rm J/}\psi}(q_T)$ at larger $x_F$, 
which is consistent with 
experimental data \cite{E866}.  Although we do not have all 
needed physical quantities
in Eq.~(\ref{a-JPsi}) for predicting the $\alpha_{{\rm J/}\psi}$ in
different $x_F$ regions, we can still test the universality 
of $\alpha_{{\rm J/}\psi}(q_T)$: the quadratic dependence on $q_T$.
In Fig.~\ref{fig3}, we plot our fits using 
$\alpha_{{\rm J/}\psi}(q_T)$ in Eq.~(\ref{a-JPsi}) and compared it 
with  E866 data in three $x_F$ regions: 
small (SXF), intermediate (IXF), and large (LXF).
Clearly, our universal functional form 
for $\alpha_{{\rm J/}\psi}(q_T)$
is consistent with all data in small $q_T$ region ($q_T <
q_T^L \sim M_{{\rm J/}\psi}/2$). 

In summary, we derived an universal functional form for  
$\alpha(q_T)$ for both the Drell-Yan and J/$\psi$
production in small $q_T$ region ($q_T<q_T^L=\kappa Q$ with
$\kappa\sim 1/3-1/2$).  All parameters defining
$\alpha(q_T)$ in Eqs.~(\ref{alpha-qt}) and (\ref{a-JPsi}) are
completely determined by either perturbatively calculable or
independently measurable quantities. We show that
$\alpha(q_T)$ is 
extremely insensitive to the atomic weight $A$ of targets. 
For the Drell-Yan process,  $\alpha(q_T)$  
in Eq.~(\ref{alpha-qt})  can be naturally connected to the
perturbatively calculated $\alpha(q_T)$ at large $q_T$ \cite{Guo1}.  
A similar test can also be carried out
for J/$\psi$ production.  J/$\psi$ suppression in
relativistic heavy ion collisions was predicted to signal the color
deconfinement \cite{Satz}.  On the other hand, significant J/$\psi$ 
suppression has been observed in  
hadron-nucleus collisions \cite{JPsi-sup}. Therefore, 
understanding the features observed in $\alpha(q_T)$ for 
J/$\psi$ production is very valuable
for finding the true mechanism of J/$\psi$ suppression.

We thank M.J. Leitch, J.M. Moss, and J.-C. Peng for helpful 
communications about experiments.  This work was supported in
part by the U.S. Department of Energy under Grant Nos.
DE-FG02-87ER40731 and  DE-FG02-96ER40989.

%%%%%%%%%%%%%% Begin References %%%%%%%%%%%%%%%%%%%%%%%%%%%%%%%%%%%%

%%%%%%%%%%%%%% End References %%%%%%%%%%%%%%%%%%%%%%%%%%%%%%%%%%%%%%%%
%%%%%%%%%%%%%% Begin Figure Captions %%%%%%%%%%%%%%%%%%%%%%%%%%%%%%%%%%%
%%%%%%%%%%%%%% End of Figure Captions %%%%%%%%%%%%%%%%%%%%%%%%%%%%%%%%%%

\end{document}